\documentclass[10pt,conference]{IEEEtran}
\IEEEoverridecommandlockouts
\usepackage[dvipsnames]{xcolor}
\usepackage{tikz}
\usepackage{amsmath}
\usepackage{multirow}
\usepackage{multicol}
\usepackage{amsfonts}
\usepackage{hyperref}
\usepackage{booktabs}
\usepackage{filecontents}
\usepackage{balance}
\usepackage{xspace}
\usepackage{soul}
\usepackage{subfigure}
\usepackage{enumitem}
\usepackage[most]{tcolorbox}
\usepackage[ruled,linesnumbered]{algorithm2e}
\usepackage[]{collab}
\usepackage{makecell}

\usepackage{url}

\SetAlCapNameFnt{\footnotesize}
\SetAlCapFnt{\footnotesize}
\SetKwInput{kwInit}{Init}

\definecolor{ballblue}{rgb}{0.13, 0.67, 0.8}
\collabAuthor{jy}{ballblue}{Jinyang Liu}
\collabAuthor{zb}{teal}{Zhuangbin Chen}
\collabAuthor{ky}{purple}{Yu Kang}
\collabAuthor{slhe}{blue}{ShiLin He}
\collabAuthor{llq}{yellow!55!red}{Liqun Li}
\collabAuthor{hy}{red}{Hongyu Zhang}
\collabAuthor{pj}{orange}{Pinjia He}

\def\BibTeX{{\rm B\kern-.05em{\sc i\kern-.025em b}\kern-.08em
    T\kern-.1667em\lower.7ex\hbox{E}\kern-.125emX}}

\newcommand\cloud{Azure\xspace}
\newcommand\nm{iPACK\xspace}
\newcommand\pf{GIP\xspace}

\begin{document}
\author{\large Jinyang Liu\IEEEauthorrefmark{1}, Shilin He\IEEEauthorrefmark{2}, Zhuangbin Chen\IEEEauthorrefmark{1}, Liqun Li\IEEEauthorrefmark{2}, Yu Kang\IEEEauthorrefmark{2}, Xu Zhang\IEEEauthorrefmark{2}, Pinjia He\IEEEauthorrefmark{3}, \\ Hongyu Zhang\IEEEauthorrefmark{4}, Qingwei Lin\IEEEauthorrefmark{2}\IEEEauthorrefmark{7}\thanks{\hspace{-2ex}\IEEEauthorrefmark{7}Qingwei Lin is the corresponding author.}, Zhangwei Xu\IEEEauthorrefmark{5}, Saravan Rajmohan\IEEEauthorrefmark{6}, Dongmei Zhang\IEEEauthorrefmark{2}, Michael R. Lyu\IEEEauthorrefmark{1}\\
\IEEEauthorblockA{
\IEEEauthorrefmark{1}The Chinese University of Hong Kong, Hong Kong SAR, China, \{jyliu, zbchen, lyu\}@cse.cuhk.edu.hk\\
\IEEEauthorrefmark{2}Microsoft Research, Beijing, China, \{shilin.he, liqli, kay, xuzhang2, qlin, dongmeiz\}@microsoft.com\\
\IEEEauthorrefmark{3}The Chinese University of Hong Kong, Shenzhen, China, hepinjia@cuhk.edu.cn\\
\IEEEauthorrefmark{4}Chongqing University, Chongqing, China, hyzhang@cqu.edu.cn\\
\IEEEauthorrefmark{5}Microsoft Azure, Redmond, USA, zhangxu@microsoft.com\\
\IEEEauthorrefmark{6}Microsoft 365, Redmond, USA, saravanakumarrajmohan@outlook.com
}
}

%
  
\title{
Incident-aware Duplicate Ticket Aggregation\\for Cloud Systems
}

\pagestyle{plain}

\maketitle

\begin{abstract}
In cloud systems, incidents are potential threats to customer satisfaction and business revenue. When customers are affected by incidents, they often request customer support service (CSS) from the cloud provider by submitting a support ticket.
Many tickets could be duplicate as they
are reported in a distributed and uncoordinated manner.
Thus, aggregating such duplicate tickets is essential for efficient ticket management. Previous studies mainly rely on tickets' textual similarity to detect duplication; however, duplicate tickets in a cloud system could carry semantically different descriptions due to the complex service dependency of the cloud system. To tackle this problem, we propose \nm, an incident-aware method for aggregating duplicate tickets by fusing the failure information between the customer side (i.e., tickets) and the cloud side (i.e., incidents).
We extensively evaluate \nm on three datasets collected from the production environment of a large-scale cloud platform, \cloud.
The experimental results show that \nm can precisely and comprehensively aggregate duplicate tickets, achieving an F1 score of 0.871$\sim$0.935 and outperforming state-of-the-art methods by 12.4\%$\sim$31.2\%.
\end{abstract}


\section{Introduction}
In the era of Cloud Computing, cloud platforms such as Amazon AWS, Microsoft Azure, and Google Cloud Platform serve millions of users worldwide.
When customers 
encounter a technical problem with the platform; they often resort to cloud providers for help by submitting a \textit{support ticket} (ticket for short), which consists of a textual issue description and some basic attributes (e.g., date and product name).
From the cloud provider's perspective, once a ticket is received, it is essential to provide timely assistance to customers to avoid user dissatisfaction and financial loss~\cite{aws_response}\cite{azure_response}.  

In practice, \textit{incidents} (i.e., unexpected service interruptions) are inevitable for large-scale cloud platforms~\cite{liu2019bugs}\cite{cotroneo2019bad}.
Though much effort has been devoted to ensure the reliability of cloud systems~\cite{li2021fighting}\cite{zhao2020understanding}\cite{Muhammad2021fault}, customers could still be impacted by incidents. 
For a large-scale cloud platform serving  millions of customers, incidents could trigger a large number of tickets, among which many could be duplicate as the tickets are reported in a distributed and uncoordinated manner. 
To reduce the burden of support engineers, it is essential to \emph{precisely and comprehensively aggregate the tickets, i.e., clustering the duplicate tickets caused by the same incident}.
By doing this, the support team can resolve the tickets more efficiently. 

To aggregate the tickets caused by the same incident, a common practice 
is to check if multiple tickets with similar symptom descriptions are reported within a short period.
The intuition behind this is that customers using the same functionalities or services tend to 
encounter similar problems if they are caused by the same incident (e.g., service unavailability).
Most existing studies on duplicate issue report detection measure the semantic similarity between two reports based on their textual descriptions, using natural language processing techniques such as word frequency~\cite{sun2010discriminative}\cite{sun2011towards}\cite{zheng2019ifeedback}, word embedding~\cite{yang2016combining}\cite{budhiraja2018dwen}, topic modeling~\cite{budhiraja2018lwe}, and  pretrained model~\cite{haering2021automatically}.
Such semantic similarity-based approaches work well for traditional software systems 
(e.g., NetBeans~\cite{lazar2014generating}, Eclipse and Firefox~\cite{lamkanfi2013eclipse}).
However, they are sub-optimal for aggregating duplicate tickets in cloud systems due to the large-scale and heterogeneous architecture of cloud systems~\cite{cotroneo2019bad}\cite{yang2021aid}\cite{wang2021fast}. 
The main reason is that customers of cloud systems could encounter distinct issues (with distinct symptoms) caused by the same incident.
On the one hand, customers using the same service may experience different issues due to various usage scenarios.
For example, when the control plane of the virtual machine (VM) service is problematic, the customers could complain about various problems related to VM creation, upgrade or deletion, depending on their particular scenarios.
On the other hand, multiple services can be impacted by the same incident due to the notorious failure propagation problem~\cite{li2021fighting}\cite{wang2021fast}\cite{chen2021graph} in cloud systems.
For example, when an infrastructure-level service (e.g., a storage service) is interrupted, other services depending on it (e.g., VM and Web application) can be impacted too. 
As a result, customers using different services may observe different symptoms and submit tickets with dissimilar descriptions. 
Consequently, it is insufficient to tackle this problem by solely utilizing textual descriptions of tickets. 



To address existing studies' limitations, we propose introducing cloud-side runtime information, i.e., \textit{alerts}, to facilitate ticket aggregation in cloud systems.
Modern cloud systems widely adopt monitors to continuously detect anomalies (unexpected behaviors) of cloud systems~\cite{aws_cloudwatch}\cite{azure_monitor}\cite{gcp_alerting}.
Once an anomaly is detected, an alert describing the anomaly will be fired to notify on-call engineers for inspection promptly.
The services (and their internal components) are interdependent in cloud systems~\cite{yang2021aid}\cite{wang2021fast}; therefore, when an incident impacts multiple components or services, multiple alerts will be triggered within a short period~\cite{zhao2020understanding}\cite{chen2020towards}, that is, these alerts are correlated with each other (i.e., \textbf{alert-alert relation}).
According to our study in \cloud, the correlated alerts caused by most (93\%) incidents are fired within four hours.
On the other hand, a particular issue of a component (e.g., problematic API for VM allocation) in cloud systems can reflect a particular customer-side issue (e.g., cannot create a VM).
So, it is possible to find a \textit{responsible alert} within the component that captures the issue resulting in the ticket (i.e., \textbf{ticket-alert relation}). 
In \cloud, we find that for 92\% of customer tickets; the alert system has already fired responsible alerts that cause these tickets before the tickets are submitted.

Motivated by these two kinds of relations, we propose to formulate the ticket aggregation problem in cloud systems as a two-stage linking problem, i.e., alert-alert linking and ticket-alert linking.
Intuitively, if the same incident triggers multiple inter-linked alerts and these alerts are further linked to different tickets, then we consider these tickets should be aggregated (i.e., caused by the same incident).
In doing this, it is possible to aggregate semantically different tickets via alert-alert links.

However, designing such a framework mainly faces two challenges originating from the large scale and complexity of cloud systems:
First, alerts are massive and noisy.
The main reason is that cloud systems consist of a large number of interdependent services.
Each service adopts comprehensive monitors to capture any abnormal patterns to ensure its reliability~\cite{chen2022online}. These monitors could be sensitive.
As a result, various alerts are continuously fired every second~\cite{li2021fighting}, so it is challenging to correctly identify and link alerts that are relevant to the ongoing incident. 
Second, features of both alerts and tickets have high cardinality, which means each of their features consists of too many unique values. 
When considering linking alerts and tickets, the number of feature combinations grows exponentially due to the high cardinality.
Consequently, it is hard to identify effective feature combinations between them and conduct correct correlation.


In this paper, we propose \nm to address these challenges. Specifically, \nm mainly consists of three steps, i.e., \textit{alert parsing}, \textit{incident profiling} and \textit{ticket-event correlation}.
The first two steps address the first challenge, and the third step addresses the second challenge.
In the \textit{alert parsing} step, we preprocess (parse) alerts as more coarse-grained \textit{events} to reduce redundant alerts.
Next, in the \textit{incident profiling} step, we propose GIP (graph-based incident profiling) to automatically filter noisy events and link events caused by the same incident.
As a result, each incident is represented as an event graph by considering alert-alert relations.
Afterward, in the \textit{ticket-event correlation} step, we propose AIN (attentive interaction network) to correlate a ticket to a responsible event by considering ticket-alert relations.
Finally, we aggregate these tickets that are linked to the events within the same event graph (i.e., incident), which are provided to CSS (customer support service) team to accelerate processing the tickets.

This work makes the following major contributions:
\begin{itemize}[leftmargin=*, topsep=0pt]
    
    \item We are the first to propose to introduce cloud runtime information (i.e., alerts) to aggregate duplicate tickets. 
    We propose \nm to leverage the alert-alert relations and ticket-alert relations to achieve this goal.
    
    \item We evaluate \nm on three datasets collected from the production environment of \cloud.
    The evaluation results show that \nm outperforms state-of-the-art methods by 12.4\%$\sim$31.2\%, which confirm the effectiveness of \nm.
    We also share our industrial experience of applying \nm in a large-scale cloud platform, \cloud.
\end{itemize}

\section{Background and Motivating Example}
\subsection{Background}
\label{sec: bg_alert_and_tickets}

\paragraph{\textbf{Alert}}

\begin{figure}[t]
    \centering
    \includegraphics[width=0.8\columnwidth]{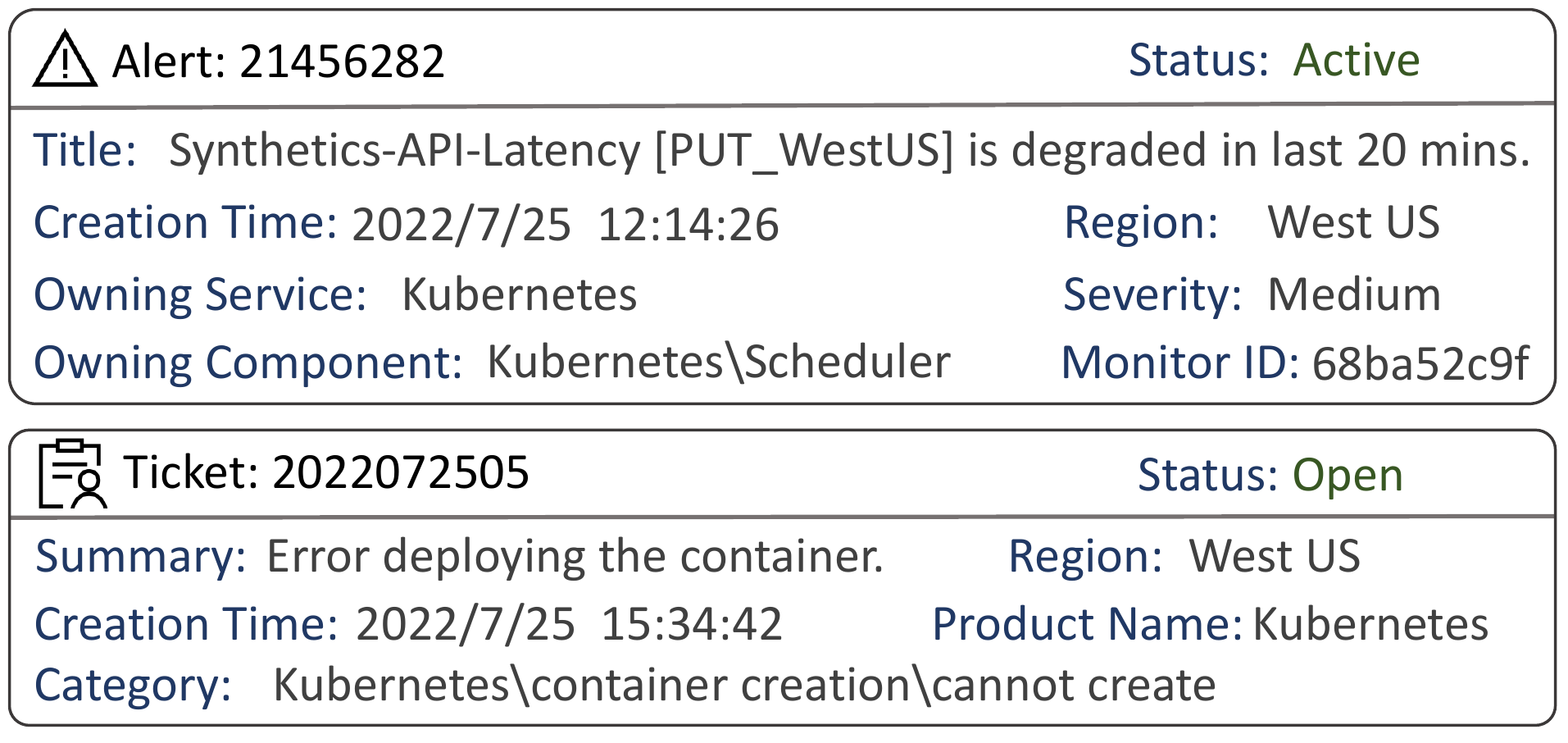}
    \vspace{-2ex}
    \caption{An example of an alert and its resultant ticket.}
    \label{fig: bg_cases}
\end{figure}

Alerts are fired by monitors that continuously detect anomalies in cloud systems, which automatically notify on-call engineers for investigation~\cite{yang2021aid}\cite{yang2022characterizing}\cite{chen2020towards}.
An alert has many attributes as presented in Fig.~\ref{fig: bg_cases} (top), including \textit{alert ID}, \textit{title}, \textit{creation time}, \textit{region}, \textit{owning service}, \textit{owning component}, \textit{severity}, \textit{monitor ID}, etc. 
The \textit{title} is generated by following a template pre-defined by engineers. 
The \textit{severity} indicates how serious the issue is, which has three levels, i.e., low, medium and high. 
A service (\textit{owning service}) consists of many components (\textit{owning component}), where each component has its own functionality or feature. 

\paragraph{\textbf{Support Ticket}}

As presented in Fig.~\ref{fig: bg_cases} (bottom), a support ticket usually contains attributes such as \textit{ticket ID}, \textit{creation time}, \textit{summary}, \textit{region}, \textit{product name}, and \textit{category}. The \textit{summary} is free text written by customers in natural language. 
The \textit{region} is where the customer's product is deployed. 
The \textit{category} is a coarse-to-fine text description initially selected by the customer, which facilitates triaging a ticket to a proper support engineer.
In addition, a ticket may also include a long detailed description (hidden in the figure).

Modern cloud platforms adopt similar schemes of alerts and support tickets described above.
For example, CloudWatch of AWS~\cite{aws_cloudwatch}, Alerting of GCP~\cite{gcp_alerting} and Azure Monitor~\cite{azure_monitor} share similar alerting mechanisms, and their alerts carry similar attributes.
Besides, their ticket management systems require similar attributes from customers as in Fig.~\ref{fig: bg_cases}, i.e., AWS Support~\cite{aws_ticket}, Google Support Hub~\cite{gcp_ticket}, and Azure Support~\cite{gcp_ticket}.
In this work, we only leverage the \textit{common} features that all these popular cloud platforms own to ensure generalizability.

\subsection{Alert-Alert Relation}\label{sec: alert_alert_relation}
\begin{figure}[t]
  \centering
  \mbox{
     \subfigure[The number of services impacted by incidents \label{fig: cross-service-alerts}]{\includegraphics[width=0.465\columnwidth]{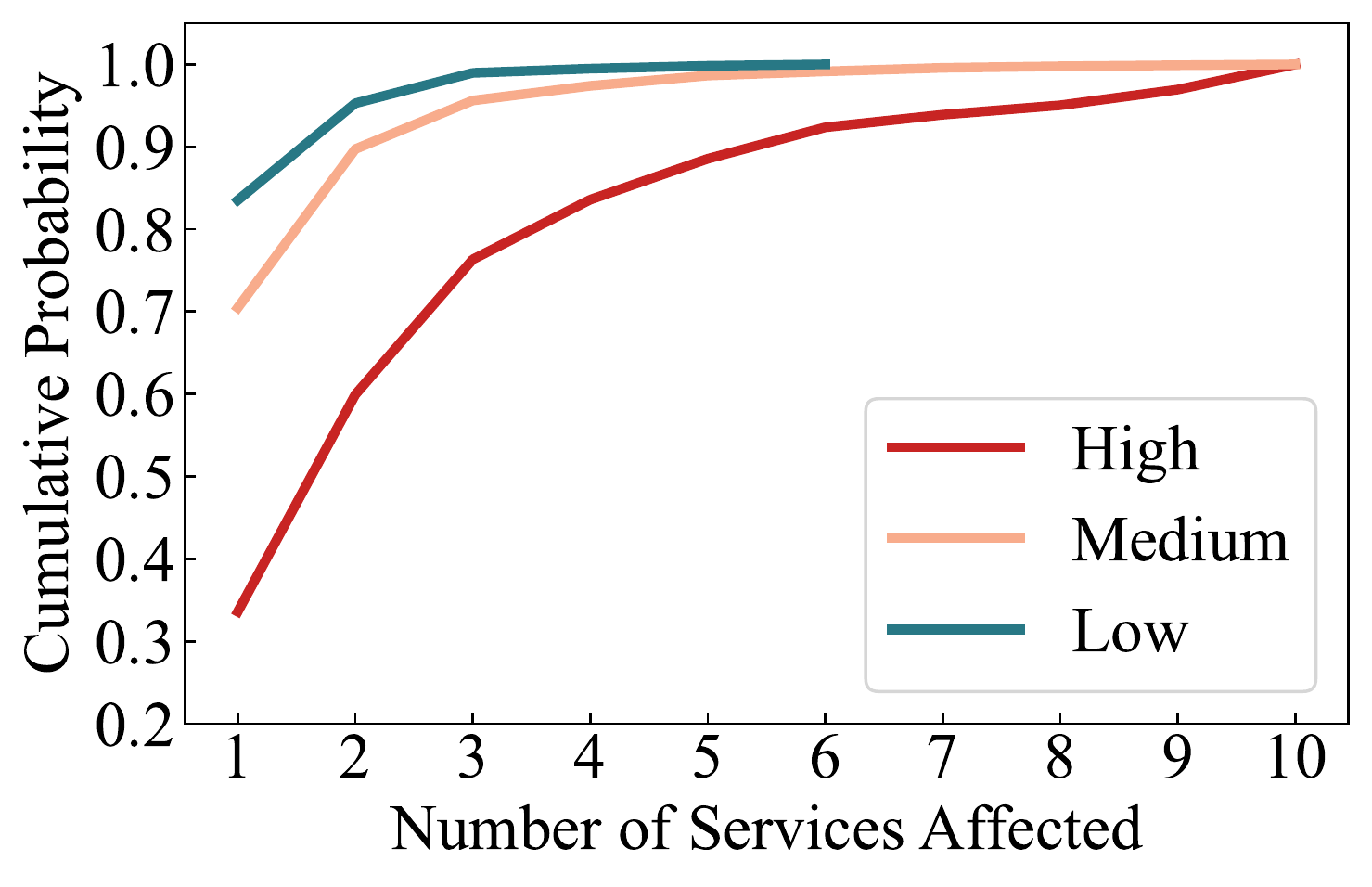}}\quad
    
    \subfigure[Distribution of max alert duration of incidents
    \label{fig: incident_duration}]
    {\includegraphics[width=0.465\columnwidth]{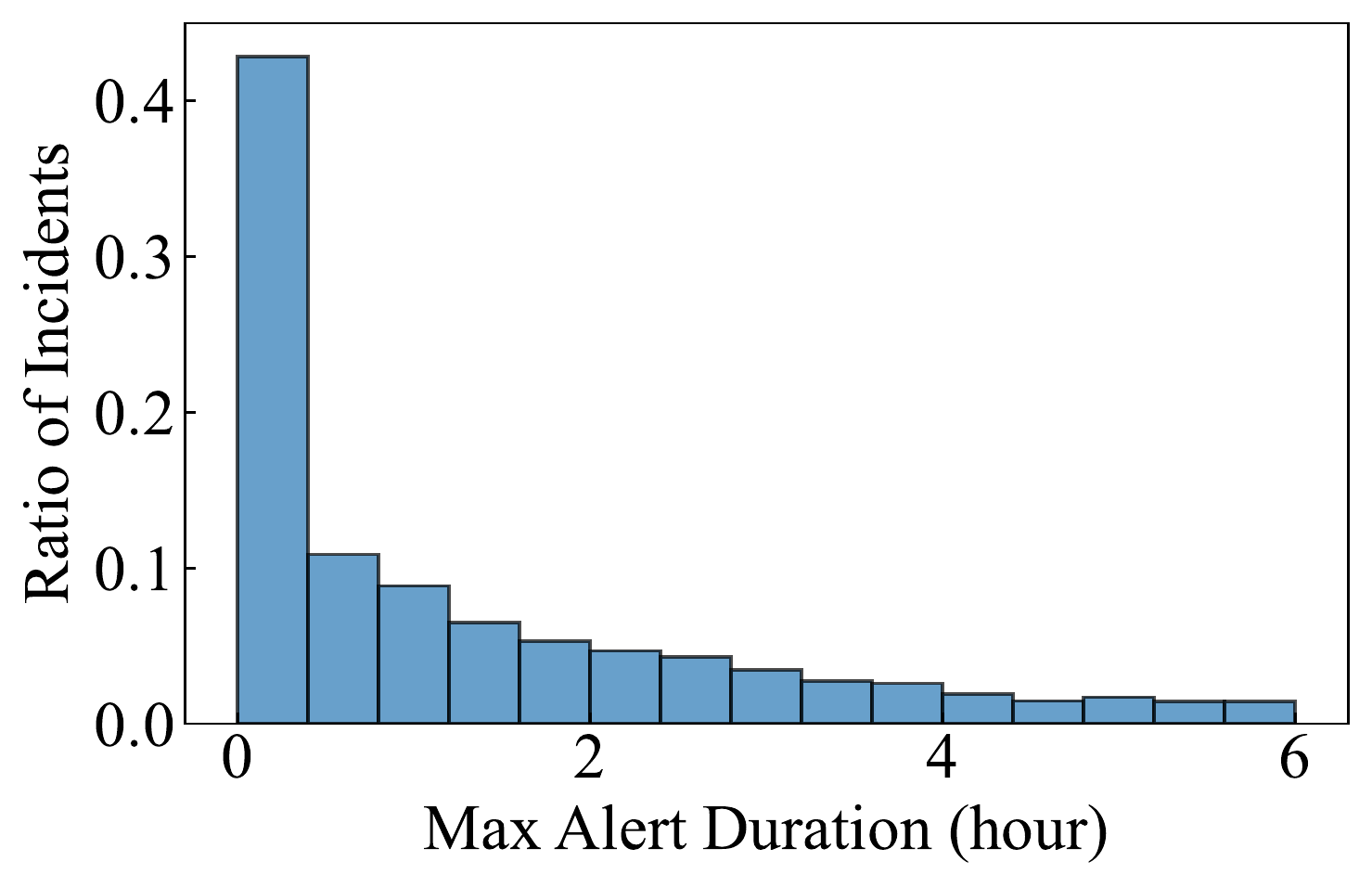}}
    }
  \vspace{-3ex}
  \caption{Statistics of alert and incident data in \cloud.}
  \label{main figure label}
\end{figure}

The alert-alert relation denotes that two alerts could be correlated if they are caused by the same incident.
The relation originates from the hierarchical structures of modern cloud systems that consist of inter-dependent components or services~\cite{chen2022online}.
When an incident happens, multiple components or services could be impacted due to failure propagation~\cite{wang2021fast}\cite{chen2020towards}, which will fire alerts within a short period associated with the same incident.
During the diagnosis of an incident, in \cloud, on-call engineers will manually mark these alerts and assess the severity of the incident according to the number of customers impacted.
According to the diagnosis history in \cloud from 2020/01/01 to 2022/06/01, as shown in Fig.~\ref{fig: cross-service-alerts}, we found incidents with a higher severity tend to affect more services. 
Especially, 70\% of high-severity incidents affect more than one service.
We studied the resultant alerts of historical incidents. We calculated the max alert duration of the incidents (i.e., the time interval between the earliest and the latest alerts triggered by the incident).
As shown in Fig.~\ref{fig: incident_duration}, we found that the max alert duration of 93\% of incidents is within four hours.
This serves as evidence to automatically identify the correlated alerts within an incident~(in Section~\ref{sec: incident_profiling}).

\begin{figure}[t]
\centering
\begin{minipage}{.465\columnwidth}
  \centering
  \includegraphics[width=\columnwidth]{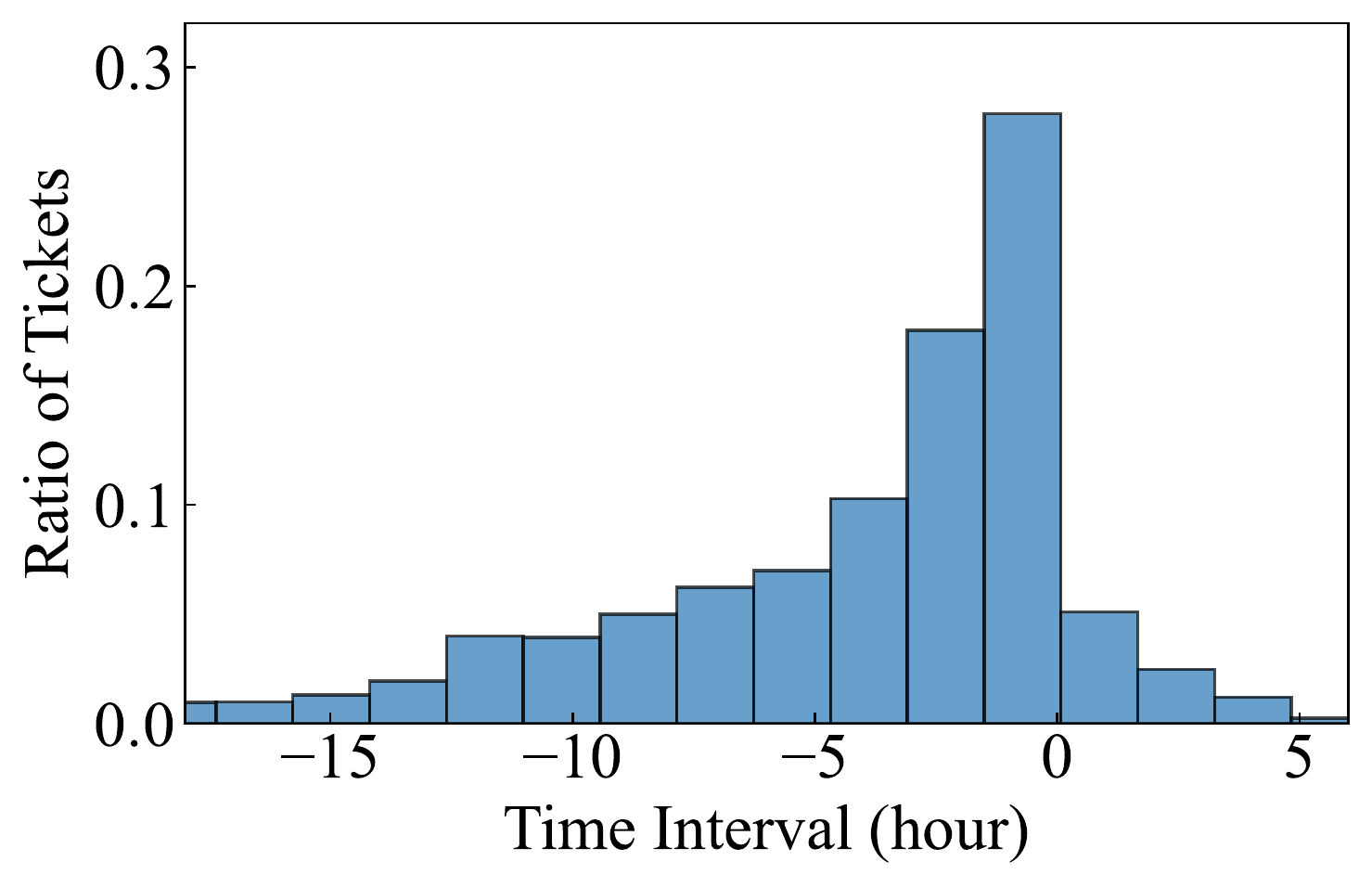}
  \vspace{-4ex}
  \caption{Time interval between alerts and resultant tickets}
  \label{fig: time_interval}
\end{minipage}\quad%
\begin{minipage}{.465\columnwidth}
  \centering
  \includegraphics[width=\columnwidth]{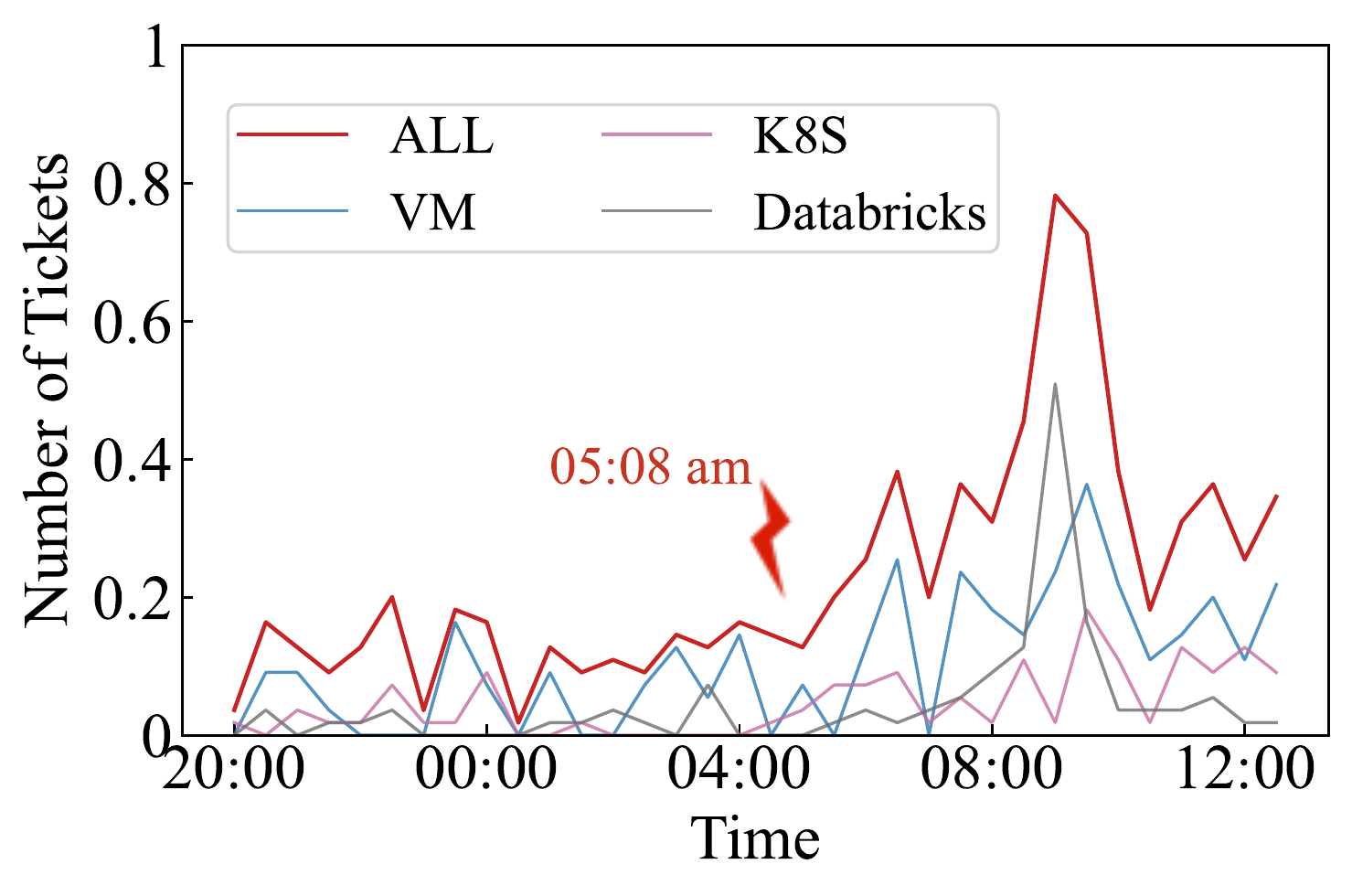}
  \vspace{-4ex}
  \caption{Ticket number trend during an incident}
  \label{fig: ticket_num_change}
\end{minipage}
\end{figure}
\subsection{Ticket-Alert Relation}\label{sec: alert_ticket_relation}
The ticket-alert relation denotes that a ticket can correlate with a responsible alert inside the cloud systems.
When a particular type of issue happens inside the cloud system (alerts are also fired), the customer could experience particular problems.
Fig.~\ref{fig: bg_cases} presents an example. If the API PUT (for container allocation) in the Kubernetes services is degraded, the customer can experience an error when deploying a container.
In \cloud, if a ticket is related to a cloud-side issue, the support engineers are required to annotate the responsible alert ID after diagnosis.
Based on the annotated alert-ticket pairs collected from 2020/01/01 to 2022/06/01, we study the time interval between alert generation and ticket submission. 
Fig.~\ref{fig: time_interval} shows the results, where a negative time interval indicates that an alert is fired before the ticket is submitted. We found around 92\% of tickets have responsible alerts fired before customers submit the tickets. This allows us to correlate responsible alerts for most tickets in runtime~(in Section~\ref{sec: correlation}).

For clarification, we summarize these important terminologies (i.e., alert, ticket, incident, alert-alert relation and ticket-alert relation) in Table~\ref{tab: term_definition} for reference.

\begin{table}[t]
\centering
\caption{Terminology Definition}
\label{tab: term_definition}
\begin{tabular}{c|c}
\hline
\textbf{Terminology} & \textbf{Definition} \\ \hline
Alert\centering & \makecell*[l]{An alert is triggered when abnormal behavior of\\ a component is detected.}\\ \hline
Ticket\centering & \makecell*[l]{A request raised by a customer to ask the cloud vendor\\ for help.} \\ \hline
Incident\centering & \makecell*[l]{Unexpected interruptions affecting services' availability\\ or performance, which usually trigger a series of alerts.} \\ \hline
\makecell*[c]{Alert-Alert\\Relation} & \makecell*[l]{Two alerts are correlated  if they are caused by the\\ same incident. (Section~\ref{sec: alert_alert_relation})}\\\hline
\makecell*[c]{Ticket-Alert\\Relation} & \makecell*[l]{A ticket is correlated with an alert if the former is\\ caused by the latter. (Section~\ref{sec: alert_ticket_relation})}\\
\hline
\end{tabular}
\end{table}

\subsection{A Motivating Example}\label{sec: case_analysis}
We present a real-world incident in July 2021 in \cloud and its resultant tickets as a motivating example.
The impact of the incident started at 05:08 AM (UTC). 
It was caused by the availability loss of the DiskRP (disk resource provider) service that provides a control plane service for managed disks. Since its gateway queue was full, a large proportion of incoming requests were rejected.
As a consequence, services relying on DiskRP experienced interruptions. On-call engineers' diagnosis confirmed that three services  were impacted, i.e., virtual machine (VM), Databricks, and Kubernetes (K8S).
Customers using these services were affected, which led to overwhelming tickets.
As shown in Fig.~\ref{fig: ticket_num_change}, the ticket numbers of the services simultaneously increased right after the impact started, which implies the three services could be impacted by the same incident concurrently.
In particular, the CSS team received around \textit{four} times the number of tickets than usual within a short period and assigned \textit{twice} the number of support engineers to handle these tickets.
We list some samples of alerts and tickets related to this incident in Table~\ref{tab: motivating_example}.
These tickets ($t_1 \sim t_8$) carry dissimilar semantics due to different use scenarios and services for different customers.
Therefore, it is hard to know that these tickets are actually caused by the same incident, rendering the difficulty for support engineers to group them and handle the burst of tickets efficiently.

\definecolor{aurometalsaurus}{rgb}{0.43, 0.5, 0.5}
\begin{table*}
\centering
\caption{Alerts caused by the same incident and the resultant tickets (some features are omitted due to space limitation.)}
\vspace{-2ex}
\label{tab: motivating_example}
\def\arraystretch{1.2}
\begin{tabular}{ccl|cl}
\hline
\multicolumn{1}{c|}{\multirow{2}{*}{\textbf{Service}}} &
\multicolumn{2}{c|}{\textbf{Tickets}} & \multicolumn{2}{c}{\textbf{Alerts}} \\  \cline{2-5}
\multicolumn{1}{c|}{} & \multicolumn{1}{c|}{Category } & \multicolumn{1}{c|}{Summary} & \multicolumn{1}{c|}{Component} & \multicolumn{1}{c}{Title} \\ \hline
\multicolumn{1}{c|}{\multirow{2}{*}{VM}} & \multicolumn{1}{c|}{VM/Scale Update} & $t_1$: Virtual machine scale sets \textcolor{black}{resize} issue. & \multicolumn{1}{c|}{\multirow{2}{*}{\begin{tabular}[c]{@{}c@{}}Resource\\  Provider\end{tabular}}} & \multirow{2}{*}{$a_1:$ \textcolor{black}{VMStart} Failures exceed 300 times.} \\ \cline{2-3}
\multicolumn{1}{c|}{} & \multicolumn{1}{c|}{VM/VM Start} & $t_2$: \textcolor{black}{Server} did not \textcolor{black}{start} on time. & \multicolumn{1}{c|}{} &  \\ \hline
\multicolumn{1}{c|}{\multirow{2}{*}{Databricks}} & \multicolumn{1}{c|}{Databricks/\textcolor{black}{Job Issue}} & $t_3$: Unable to \textcolor{black}{open cluster} of Databricks. & \multicolumn{1}{c|}{\multirow{2}{*}{\begin{tabular}[c]{@{}c@{}}Control \\ Plane\end{tabular}}} & \multirow{2}{*}{$a_2:$ Databricks \textcolor{black}{cluster creation} fails.} \\ \cline{2-3}
\multicolumn{1}{c|}{} & \multicolumn{1}{c|}{Databricks/Cluster Launch} & $t_4$: Unable to \textcolor{black}{provision clusters}. & \multicolumn{1}{c|}{} &  \\ \hline
\multicolumn{1}{c|}{\multirow{2}{*}{K8S}} & \multicolumn{1}{c|}{K8S/Cluster Update} & $t_5$: Unable to \textcolor{black}{autoscale}. & \multicolumn{1}{c|}{\multirow{2}{*}{\begin{tabular}[c]{@{}c@{}}Resource \\ Scheduler\end{tabular}}} & \multirow{1}{*}{$a_3:$ \textcolor{black}{The PUT operation} success rate \textless{}80\%.}  \\ \cline{2-3}
\multicolumn{1}{c|}{} & \multicolumn{1}{c|}{K8S/Cluster Update} & $t_6$: Cannot \textcolor{black}{upgrade} node pool, stuck. & \multicolumn{1}{c|}{} &  \textcolor{aurometalsaurus}{$a_4:$ CPU utilization exceeds 90\%.} \\ \hline
\end{tabular}
\vspace{-4ex}
\end{table*}

We propose to aggregate these tickets by simultaneously leveraging the aforementioned alert-alert relations and ticket-alert relations.
We take Table~\ref{tab: motivating_example} as an example to elaborate our intuition.
First, we need to know what alerts are triggered by an incident, i.e., profiling the incident. 
In this example, we link the alerts $a_1-a_2-a_3$ via capturing the alert-alert relations (i.e., they are caused by the same incident).
Second, we need to know what tickets are caused by these alerts, namely, linking $a_1-(t_1,t_2)$, $a_2-(t_3,t_4)$, and $a_3-(t_5,t_6)$.
Finally, because the alerts $a_1 \sim a_3$ are linked as an incident and $t_1 \sim t_6$ are further linked to these alerts, we can aggregate $t_1 \sim t_6$ as the same cluster even though they possess dissimilar semantics.



\noindent\textbf{Challenges.} To achieve this, \nm should address the following two challenges originated from the large scale and complicated architecture of cloud systems~\cite{li2021fighting}\cite{wang2021fast}\cite{chen2022online}.

\textit{Challenge 1: Massive and noisy alerts.}
Cloud systems could contain thousands of interdependent services.
These services are closely monitored from various aspects to capture any unexpected behaviors. 
For example, there could be hundreds, even thousands of high-severity alerts reported in \cloud per day.
Some alerts are \textbf{regular alerts} that are reported frequently (due to sensitive monitoring rules) and periodically (due to periodical monitoring). These regular alerts are generally not related to a particular cloud incident and only report usual system runtime status such as CPU/memory usage rate (e.g., $a_4$ in Table~\ref{tab: motivating_example}).
In contrast, \textbf{indicative alerts} are caused by an actual problem of cloud systems. For example, the alerts $a_1$ $\sim$ $a_3$ in Table~\ref{tab: motivating_example} are indicative alerts.
It is challenging to identify the indicative alerts and correctly link them among massive and noisy alerts.
    
\textit{Challenge 2: High feature cardinality.} High feature cardinality refers to a situation where a feature has a large number of unique values.
For example, the feature \textit{category} of a ticket has more than 3,000 options, and the features \textit{component} and \textit{monitor ID} of alerts have more than 2,000 and 10,000 options, respectively.
Using traditional one-hot encoding~\cite{li2021deeplv} methods to process these features would lead to a high-dimensional feature space, resulting in the curse of dimensionality~\cite{wiki:Curse_of_dimensionality}.
Additionally, linking alerts to tickets requires the consideration of various combinations of features between them.
However, due to the high feature cardinality, the number of possible combinations grows exponentially, making it difficult to identify the most effective combinations that accurately reflect the correlation between alerts and tickets. This constitutes a significant challenge in our work.

\section{Methodology}

\subsection{Overview of \nm}
The goal of \nm is to aggregate duplicate tickets that are caused by the same cloud incident among all tickets.
Due to the large scale and heterogeneous architecture~\cite{cotroneo2019bad}\cite{yang2021aid}\cite{wang2021fast} of cloud systems, it is 
insufficient to solely consider the textual similarity of tickets to achieve this goal.
To address this problem, we introduce cloud run-time information (i.e., alerts) and formulate it as a two-stage linking problem. Intuitively, \nm first finds links between alerts by leveraging alert-alert relations. 
These inter-linked alerts constitute a graph to represent an incident.
Then \nm identifies the tickets that are caused by these alerts according to ticket-alert relations.
The tickets linked to the alerts within the same graph (i.e., incident) are aggregated.
Thus, we can aggregate the tickets with dissimilar semantics via the bridge of alert-alert links.

\begin{figure}[t]
    \centering
    \includegraphics[width=1\linewidth]{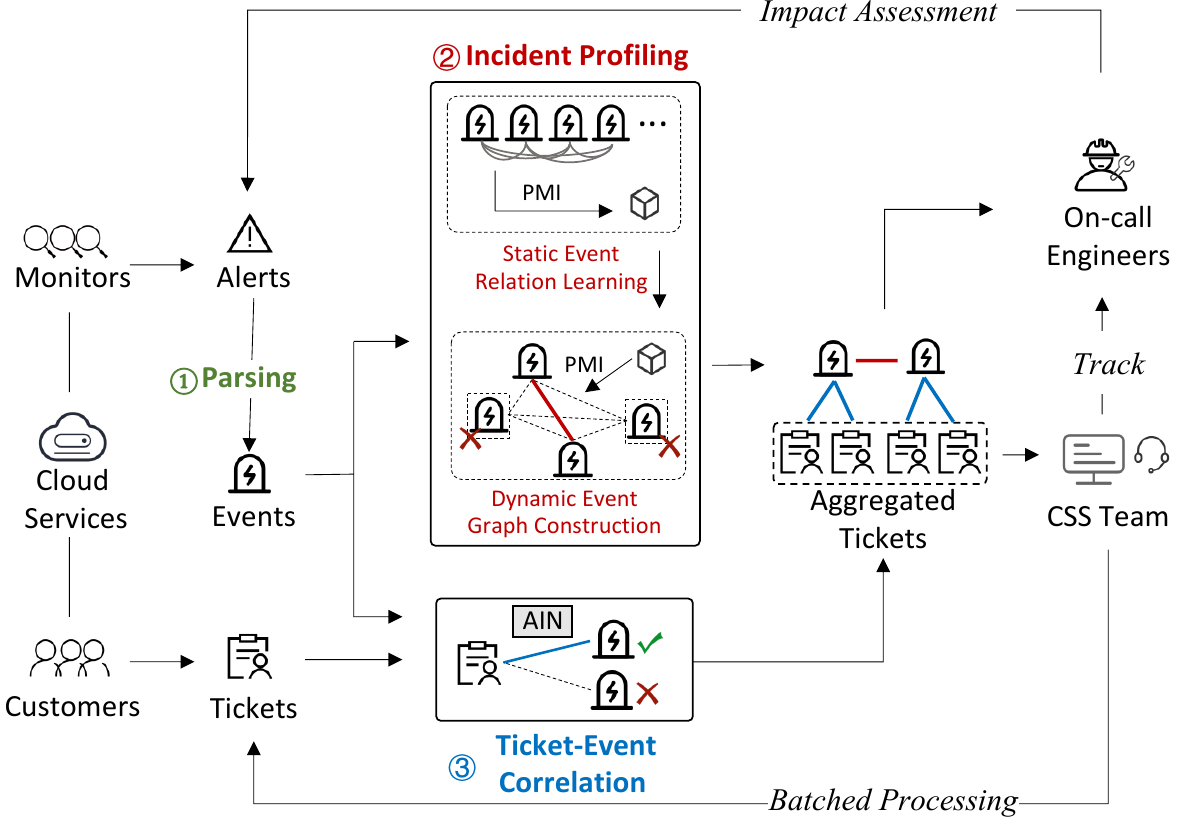}
    \caption{The overall framework of \nm.} 
    \label{fig: overall_framework}
\end{figure}

As shown in Fig.~\ref{fig: overall_framework}, \nm consists of three steps: \textit{alert parsing}, \textit{incident profiling} and \textit{ticket-event correlation}. In the \textit{alert parsing} step, we parse alerts as more coarse-grained \textit{events} to reduce redundant alerts.
Next, in the \textit{incident profiling step}, we propose a graph-based incident profiling (GIP) method to remove the regular events (i.e., parsed regular alerts) and link correlated indicative events.
Then, in the \textit{ticket-event correlation}, we propose an attentive interaction network (AIN) to correlate a ticket to an event.
Finally, if two tickets are correlated to the events within the same event graph (i.e., the same incident), we aggregate the tickets as the same cluster. 
The results of the ticket aggregation are presented to the CSS (Customer Support Services) team to streamline the ticket processing process and improve efficiency. This allows support engineers to send out batch notifications to potentially affected customers and provide quick guidance for service recovery. Additionally, the results can aid on-call engineers in conducting impact assessments, including identifying affected services and determining the extent of customer impact caused by the incident (e.g., number of affected customers).


\subsection{Alert Parsing}\label{sec: alert_parsing} 
The title of an alert is generated following an engineer-specified template.
Monitors may be triggered multiple times during an incident causing massive redundant alerts.
To reduce the volume of alerts and avoid redundancy, we parse each alert to its corresponding template and aggregate the alerts sharing the same template as an \textbf{event}.
Take $a_1$ in Table~\ref{tab: motivating_example} as an example; multiple similar alerts can fire concurrently such as ``VMStart Failures exceed 100/150/200/250 times'', which are aggregated as ``VMStart Failures exceed~$\ast$~times''.

We formulate this problem as the well-studied log parsing problem~\cite{zhu2019tools} following~\cite{wang2021fast}. We propose to customize a widely-adopted log parsing algorithm,  Drain~\cite{he2017drain} to parse the alerts into templates (events). 
Drain works by extracting the common parts of alert titles from each group of alerts, where the group is determined by calculating the overlap of words. 
To enhance Drain in our scenario, we observe that if two alerts are reported by different monitors or belong to different components, the two alerts must have distinct templates.
Therefore, we first divide all alerts into different partitions according to both \textit{monitor~ID} and \textit{owning component}. We then apply Drain in each partition to extract the templates. 
In this way, we can reduce the noises in each partition and also accelerate the processing by parallel computing.
Finally, each alert is parsed as an event, which introduces two features, i.e., \textit{event template} and \textit{event~ID} (a hash value of its template).
Within a fixed time window (Section~\ref{sec: incident_profiling}), for events sharing the same template, we reserve the latest event and discard the rest of the events to reduce its volume.
The following steps are applied to events instead of raw alerts.

\subsection{Incident Profiling}\label{sec: incident_profiling}
The goal of this step is to represent an incident via \textit{linking the correlated events that are caused by the same incidents}. 
In doing this, the linked events can then be used to bridge semantically different tickets in the next step (Section~\ref{sec: correlation}).

To learn relations between events, some existing solutions leverage manual annotations~\cite{chen2022online}\cite{chen2020identifying}, which are not practical because such labels are hard to obtain and usually insufficient in real-world practice. 
While there are unsupervised solutions~\cite{zhao2020understanding}\cite{chen2021graph}, they require prior knowledge (e.g., the precise topology of cloud services) to estimate alert relations. However, such prior knowledge is usually inaccurate and requires extensive efforts to collect, update and validate~\cite{yang2021aid}\cite{chen2021graph}\cite{chen2022online}.

We propose an unsupervised approach, i.e., \underline{G}raph-based \underline{I}ncident \underline{P}rofiling (\pf), which does not rely on prior knowledge. 
The input is a series of events within a time window, and the output is one or multiple graphs of the events. 
Each graph profiles an incident containing indicative events related to the incident. 
GIP has a \textit{static event relation learning} step and a \textit{dynamic event graph construction} step.
Intuitively, if two events are correlated, these events are more likely to be triggered within a short period frequently in the history~\cite{chen2021graph}\cite{chen2022online}.
We model such frequent patterns in the first \textit{static event relation learning} step.
Then, in the \textit{dynamic event graph construction} step, we dynamically link the events possessing the learned frequent patterns and remove regular events in the runtime.



\subsubsection{Static Event Relation Learning}\label{sec: static_event_relation_learning}
In this step, we assign a static score to each event pair weighing how likely they co-occur in history. 
To this end, we first collect a series of historical events in chronological order.
Then we apply a \textit{four-hour-long} sliding time window on these events with a step size of one hour. We adopt \textit{four} hours as the window length because it can cover most alerts within an incident according to our study in Section~\ref{sec: alert_alert_relation}. The one-hour step size allows us to introduce enough new events for learning the static event relations and avoid separating co-occurred events into two different windows.
Each window $w_i$ contain multiple events, i.e., $w_i = [e_1, e_2, e_3, ... ]$.
If two events appear in the same window, we count it as a co-occurrence. 
Based on these windows, we compute the point-wise mutual information (PMI) score~\cite{wiki:Pointwise_mutual_information} for each event pair, which is a popular metric to measure co-occurrence associations~\cite{qin2021relation}\cite{yao2019graph}.
Formally, the PMI value for the event pair $(e_i, e_j)$ is :
\begin{align}
    PMI(e_i, e_j) = log\frac{p(e_i,e_j)}{p(e_i) p(e_j)},
\end{align}
where $p(e_i,e_j) = \frac{C(e_i, e_j)}{M}$,  $p(e_i) = \frac{C(e_i)}{M}$. $C(e_i, e_j)$ denotes the number of windows that contain both $e_i$ and $e_j$, and $C(e_i)$ is the number of windows that contain $e_i$. $M$ is the total number of windows.
A higher PMI value indicates two events are more likely to co-occur in history, and a positive PMI value indicates they are more likely to co-occur than appear individually.
We use $d(e_i, e_j)$ to denote the pre-computed PMI value for the event pair $(e_i, e_j)$.



\begin{algorithm}[t]
\small
\caption{Dynamic Event Graph Construction}
\label{algo: graph_construction}
\SetAlgoLined
\KwIn{Pre-computed PMI values in $d$, a window of latest events $w_j=[e_1, e_2, e_3, ...]$, hyper-parameter $\mu\in$[0,1]}
\KwOut{$g_o=\{g_1, g_2, ...\}$}

\kwInit{g $\leftarrow$ Empty undirected graph; 
r $\leftarrow$ Empty list
}

\For{$i\leftarrow 1$ \KwTo $l$}{
    \For{$j\leftarrow i$ \KwTo $l$}{
        \If{
        $d(e_i, e_j) > 0$
        }{
        
        g.AddWeightedEdge(($e_i$, $e_j$),
        weight=$d$($e_i$, $e_j$))
        }

        }    
}

\For{each node $e_i \in$ g}{
    $\mathcal{W}$ = GetWeightsOfOutEdges($e_i$)
    
    AscendingSort($\mathcal{W}$)
    
    $\gamma$ = SearchKneePoint($\mathcal{W}$) // Kneedle algorithm
    
    \If{$\gamma< \mu$} {
        g.RemoveNode($e_i$)
    }
}
$g_o \leftarrow$ GetSubGraphs($g$)
\end{algorithm}


    
    


\subsubsection{Dynamic Event Graph Construction}
We then dynamically construct event graphs in the runtime by utilizing the learned static PMI values.
The input to this step is the events collected within the latest four-hour-long time window.
The output is one or more event graphs, each of which contains correlated events caused by the incident.

Intuitively, we aim to link the events with high PMI values because they are possibly caused by the same ongoing incident in the runtime, considering they frequently co-occur in history.
However, regular (noisy) events tend to co-occur with various types of events because they frequently appear regardless of whether there is an incident.
In contrast, indicative events only frequently co-occur with only a small portion of events.
Based on the difference between regular events and indicative events, we propose a novel algorithm to prune the regular events automatically, and the remaining indicative events are correlated.
The pseudocode of the algorithm is shown in Algorithm~\ref{algo: graph_construction}.
First, we link every pair of events with positive PMI values constituting a single initial event graph $g$ with the PMI values as weights of edges (line $1\sim 7$).
Then, for each node, we calculate a knee point (i.e., $\gamma$ in Algorithm~\ref{algo: graph_construction}) based on the PMI values of all its out edges, i.e., $\mathcal{W}$ (line $9\sim 11$). 
Specifically, we adopt the Kneedle algorithm~\cite{satopaa2011finding} to calculate $\gamma$.
A small $\gamma$ for a node denotes that most PMI values of its linked neighbors are large, namely, the node frequently co-occurs with many neighbors (events). This implies that the node is more likely to be a regular event.
Therefore, we remove the node if its $\gamma$ is less than a threshold $\mu$ (line $12\sim 14$).
As revealed by previous studies~\cite{zhao2020understanding}\cite{chen2020incidental}, regular events make up a large portion of all events.
Therefore, we empirically set $\mu=0.8$ to remove most events aggressively, which turns out to be effective in our scenario (Section~\ref{sec: GIP_ablation_study}).
Finally, we extract subgraphs (i.e., connected component~\cite{wiki:Component_(graph_theory)}) from the the pruned graph $g$ (line $16$).




\subsection{Ticket-Event Correlation}\label{sec: correlation}
After profiling incidents as several event graphs (i.e., event-event linking), we correlate each ticket to the event that captures the internal cloud issue resulting in the ticket (i.e., event-ticket linking).
If two tickets are correlated to inter-linked events (i.e., they are caused by the same incident), we can then aggregate them as the same cluster.



\begin{figure}[t]
    \centering
    \includegraphics[width=1\linewidth]{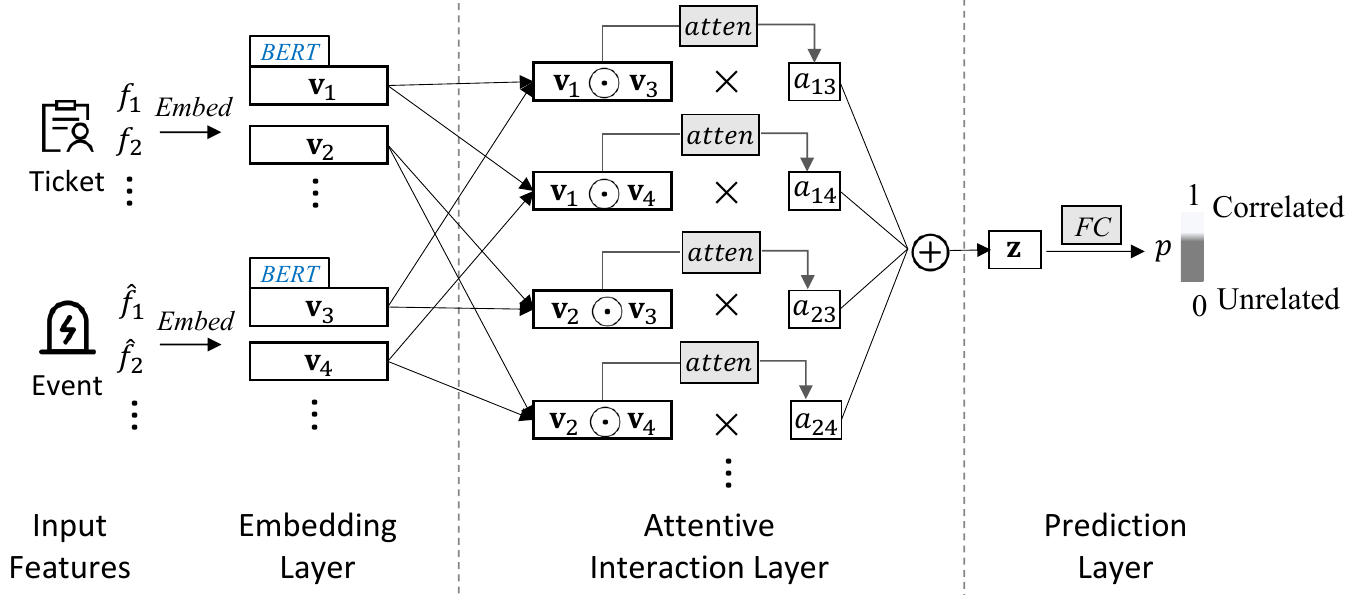}
    \caption{The overall framework of AIN.} 
    \label{fig: AIN_framework}
\end{figure}

We mainly address the challenge caused by the high cardinality of features of tickets and events (Section~\ref{sec: case_analysis}).
Inspired by factorization machine~\cite{rendle2010factorization} in the field of recommendation systems,
we propose an attentive interaction network (AIN), which decomposes feature combinations as Hadamard products of low-dimension feature embeddings.
In this way, we bypass directly encoding the exponentially-growing feature combinations with high-dimension feature vectors.
The input of AIN is a ticket-event pair and the output is a probability representing how likely the input pair is correlated.
Fig.~\ref{fig: AIN_framework} shows the overall framework of AIN composed of three layers, i.e., \textit{embedding layer}, \textit{attentive interaction layer}, and \textit{prediction layer}, which are elaborated as follows.


\textbf{Embedding Layer.} The embedding layer represents all features ($f_i$ for a ticket feature and $\hat{f}_i$ for an event feature) as trainable vectors (i.e., embeddings) denoted as $\mathbf{v_i}\in\mathbb{R}^k$, where $k$ is a user-defined hyper-parameter. 
For \textit{summary} of tickets and \textit{event template} of events (denoted as $f_1$ and $\hat{f}_1$ in Fig.~\ref{fig: AIN_framework}), we resort to the power of pretrained model BERT (Bidirectional Encoder Representations from Transformers)~\cite{DBLP:conf/naacl/DevlinCLT19} to embed their semantics as vectors. 
We exclude the detailed ticket description since it potentially introduces noises, and the summary already provides the essential part~\cite{yang2016combining}\cite{haering2021automatically}\cite{wang2008approach}.
The remaining features are initialized as random vectors.

\textbf{Attentive Interaction Layer.} 
After each feature is associated with an embedding vector, the attentive interaction layer models the feature combination of two features as the Hadamard product (i.e., element-wise product denoted as $\odot$) of their corresponding embedding vectors.
For $ \mathbf{u}=\mathbf{x}\odot \mathbf{y}$ we have $u_i=x_iy_i$. 
The attentive interaction layer models combinations of features across a ticket and an event, formally, 
\begin{align}
    \mathbf{z} = \sum_i^n\sum_j^m a_{ij} (\mathbf{v}_i\odot \mathbf{v}_j), \label{equ: afm}
\end{align}
where $n$ and $m$ are the numbers of ticket and event features, respectively. 
To identify the effective feature combinations for different ticket-event pairs, AIN computes an importance score $a_{ij}$ for each combination result $v_i\odot v_j$ in Equation~(\ref{equ: afm}).
Afterwards, these feature combinations are summarized as a single representation $\mathbf{z}\in \mathbb{R}^k$ by computing their weighted average. 
The importance weight $a_{ij}$ is calculated as follows:
\begin{align}
\hat{a}_{ij} &= \mathbf{h}^T\phi(\mathbf{W}(\mathbf{v}_i\odot\mathbf{v}_j) + \mathbf{b}), \label{equ: afm_weight}\\
    a_{ij} &= \frac{e^{\hat{a}_{ij}}}{\sum_i^n\sum_j^me^{\hat{a}_{ij}}},  \label{equ: afm_softmax}    
\end{align}
Equation~(\ref{equ: afm_weight}) denotes a fully-connected (FC) neural network that takes the combination of two features as input and outputs their (unnormalized) importance weight.
where $\phi(x)=max(0,x)$ is the ReLU activation function. $\mathbf{h}^T\in\mathbb{R}^{r}$, $\mathbf{W}\in\mathbb{R}^{(r\times k)}$ and $\mathbf{b}\in\mathbb{R}^{r}$ are trainable parameters. $r$ is a hyper-parameter that denotes the size of the hidden layer.
Equation~(\ref{equ: afm_softmax}) normalizes the importance weights to $[0,1]$.
The importance weights control how much each feature combination contributes for prediction.
For example, in Equation~(\ref{equ: afm}), for $a_{ij}$ close to~1, its corresponding feature combination will dominate the summarized vector $\mathbf{z}$. 
This means that the prediction mostly depends on the feature combination of $\mathbf{v_i}$ and $\mathbf{v_j}$. In addition, the weights are automatically learned by the FC in Equation~\ref{equ: afm_weight}, we actually force AIN to select the effective feature combinations when learning from the data.

\textbf{Prediction Layer.} 
We formulate ticket-event correlation as a binary classification problem. 
Particularly, to calculate the correlation probability $p$, an FC neural network is applied on $\mathbf{z}$, i.e., $p = \sigma(\mathbf{w}_o^T\mathbf{z} +  b_o)$,
where $\mathbf{w}_o\in\mathbb{R}^{k}$ and $b_o \in \mathbb{R}$ are trainable parameters, and $\sigma(x)={\frac {1}{1+e^{-x}}}$ is the Sigmoid function producing a probability within the range of $[0,1]$.
To update all trainable parameters, we utilize the popular Adam optimizer~\cite{diederik2015adam} to minimize the following binary cross-entropy loss $\mathcal{L}_{BCE}$ via fitting training data with $N$ ticket-event pairs. 
\begin{equation}
    \mathcal{L}_{BCE} =-\sum_{i}^{N}\big(y_i log{(p_i)} + (1-y_i)log(1-p_i)\big),
    \label{equ:bce}
\end{equation}
where $y_i=1$ for positive (i.e., correlated) ticket-event pairs and $y_i=0$ for negative (i.e., unrelated) pairs. 
The positive samples are collected by extracting the responsible alert ID of a ticket from its resolution text written by support engineers (Section~\ref{sec: alert_ticket_relation}). So, such data is gradually accumulated during the daily work routine of support engineers, which does not incur additional manual effort for data labeling.
We then randomly sample the same number of negative pairs.
The features used are \textit{event template}, \textit{event ID}, \textit{severity}, \textit{monitor ID}, \textit{owning service},  \textit{owning component} for events, and \textit{product name}, \textit{category}, \textit{summary} for tickets.

\subsection{Deployment}
\nm consists of offline parts (pre-computed) and online parts (serving online continuously) for deployment in the could systems. 
The offline parts include alert parsing, static event relations learning and AIN training.
The online parts conduct alert parsing, dynamic event graph construction, and ticket-event correlation utilizing the trained AIN.
The details are as follows.

\subsubsection{Offline Parts}
Intuitively, the offline parts leverage the historical data to prepare intermediate data (e.g., PMI values) or model (e.g., AIN) for online use. Specifically, \nm parse all collected alerts to events (Section~\ref{sec: alert_parsing}). Then, static event relations learning is conducted (Section~\ref{sec: static_event_relation_learning}), which computes PMI values for all event pairs. The PMI values are then stored in a Redis database for reference.
After that, AIN is trained using historical ticket-event pairs. 
\cloud continuously collects the alert and ticket data; in order to capture the latest system update (e.g., new alerts), the offline parts are executed periodically (e.g., once every month).

\subsubsection{Online Parts}
In the online deployment, \nm is periodically executed (e.g., every five minutes) and pushes its latest analysis results to the CSS team. Support engineers can also manually trigger \nm when needed (e.g., a large volume of tickets are received).
Considering cloud services and customers are physically isolated in different regions, \nm is applied separately in different regions.
Once executed, \nm collects the latest alerts and tickets within the latest four-hour-long time window to analyze. We can reduce the great volume of ticket-event pairs by filtering with region and time.
The tickets and alerts in the same time window and region constitute a \emph{chunk}.

In each chunk, after parsing alerts as events, GIP is applied to link events as event graphs (i.e., incidents).
Then, we apply AIN to link each ticket to one of the events. 
For each ticket, AIN recommends a list of events ranked by the associated correlation probabilities.
Note that we exclude the tickets whose largest probability in the ranked list is smaller than a confidence threshold $\theta=0.8$, because they are more likely caused by a customer-side issue (e.g., incorrect configurations).
Next, tickets that are correlated to the events within the same event graph are aggregated as a cluster.
Based on the aggregation results, on the one hand, on-call engineers can conduct impact assessment (i.e., how many customers are impacted) for an incident; on the other hand, the CSS team can avoid duplicate manual inspection and make batched communication to customers. (e.g., provide the latest mitigation progress of the internal incident).


\section{Experiments}

We answer the following research questions (RQs) to evaluate the performance of \nm:

\begin{itemize}[leftmargin=*, topsep=0pt]
    \item \textbf{RQ1}: How effective is \nm in aggregating duplicate tickets caused by the same incident?
    \item \textbf{RQ2}: How effective is AIN in correlating a ticket to the responsible event?
    \item \textbf{RQ3}: How does graph-based incident profiling (GIP) impact the effectiveness of \nm? 
\end{itemize}

\subsection{Experimental Setting}
\subsubsection{Dataset}\label{sec: dataset}

We collect the datasets from the production environment of \cloud from 2020/01/01 to 2022/06/01.
To evaluate the generality of \nm, we collect three datasets from different physically isolated regions (i.e., $\mathcal{A}$, $\mathcal{B}$, and $\mathcal{C}$), which cover 81 services serving different numbers of customers.
Each dataset is collected from tens of services and includes hundreds of incidents and hundreds of thousands of alerts.
For each incident, the datasets contain tens of to hundreds of resulting tickets.
Note that we hide the specific figures of the dataset statistics due to the confidential policy of \cloud.
We use the data before 2022/01/01 to compute PMI values~(Section~\ref{sec: incident_profiling}) and train AIN~(Section~\ref{sec: correlation}). The data after the date is used for evaluation.

\subsubsection{Comparative solutions}\label{sec: baselines}
Recent studies have been working on user feedback analysis such as duplicate bug report detection~\cite{wang2008approach}\cite{nguyen2012duplicate}\cite{zhou2012learning}\cite{budhiraja2018lwe}\cite{chaparro2019reformulating} and emerging issue detection~\cite{gao2018online}\cite{zheng2019ifeedback}\cite{gao2019emerging}.
We select the following state-of-the-art approaches as our comparative solutions:

\textbf{Categorization.} We aggregate tickets by referring to their feature \textit{category} (Section~\ref{sec: bg_alert_and_tickets}), i.e., if two tickets share the same category, then they are aggregated into the same cluster. 

\textbf{iFeedback.} iFeedback is proposed and adopted by WeChat in their production environment~\cite{zheng2019ifeedback}, which targets aggregating similar user feedback by identifying frequent word combinations (and groups of combinations). 
For example, if the word combination of ``pay'' and ``fail'' bursts, an issue may happen to the payment feature of the product. 


\textbf{LWE.} LWE~\cite{budhiraja2018lwe} is a method integrating Latent Dirichlet Allocation (LDA) and word embeddings to leverage the advantages of both techniques. 
LWE first utilizes LDA to represent all tickets and roughly identify candidates of duplicated tickets.
Then, the candidates are represented using word embeddings to conduct more fine-grained clustering.

\textbf{BERT.} BERT~\cite{DBLP:conf/naacl/DevlinCLT19} is a popular pretraining model in natural language processing and has shown its power in capturing the semantics of user feedback in recent studies~\cite{haering2021automatically}\cite{liu2020automated}\cite{wu2021identifying}.
Because these studies do not directly aggregate user feedback, in this work, we adopt BERT to first represent the tickets as dense vectors, based on which we use agglomerative hierarchical clustering~\cite{wiki:hierarchical_clustering} to aggregate tickets.



\begin{table*}[t]
\linespread{1.1}
\caption{Effectiveness 
of aggregating duplicate tickets caused by the same cloud incident.}
\label{tab: clustering_accuracy}
\small
\centering
\begin{tabular}{c|ccc|ccc|ccc}
\toprule
 \multirow{2}{*}{Methods} &
\multicolumn{3}{c|}{Dataset $\mathcal{A}$} & \multicolumn{3}{c|}{Dataset $\mathcal{B}$} & \multicolumn{3}{c}{Dataset $\mathcal{C}$} \\ 
& Precision & Recall & \textbf{F1 score} & Precision & Recall & \textbf{F1 score} & Precision & Recall & \textbf{F1 score}\\
  \midrule
  \midrule
Categorization & 0.930 & 0.205 & 0.336 & 0.943 & 0.373 & 0.535 & 0.925 & 0.207 & 0.338 \\
iFeedback     & 0.901 & 0.590 & \underline{0.713} & 0.876 & 0.473 & 0.614 & 0.886 & 0.626 & 0.733 \\
LWE           & 0.862 & 0.453 & 0.594 & 0.824 & 0.515 & 0.634 & 0.861 & 0.672 & \underline{0.755}\\
BERT          & 0.884 & 0.587 & 0.705 & 0.854 & 0.710 & \underline{0.775} & 0.843 & 0.629 & 0.720 \\
LinkCM        & 0.931 & 0.507 & 0.657 & 0.892 & 0.538 & 0.671 & 0.901 & 0.628 & 0.740 \\ 
\midrule
LinkCM w/ GIP   & 0.900 & 0.685 & 0.778 & 0.886 & 0.756 & 0.816 & 0.899 & 0.809 & 0.852 \\ 
\textbf{\nm}  & 0.912 & 0.960 & \textbf{0.935} & 0.882 & 0.861 & \textbf{0.871} & 0.899 & 0.888 & \textbf{0.894} \\
\bottomrule
\end{tabular}
\end{table*}

\textbf{LinkCM.} LinkCM~\cite{gu2020efficient} is proposed to facilitate the triage of a customer-reported alert by matching it with an alert of cloud systems. 
LinkCM learns the correlation by purely fusing the titles between the report and alert via a decomposable attention mechanism and transfer learning. In our scenario, if two tickets are correlated to the same event by LinkCM, they are grouped together. 
LinkCM can also link a ticket to an event as AIN does, so we combine GIP with LinkCM (i.e., \textbf{LinkCM w/ GIP}) as a strong baseline for comparison. 

\subsubsection{Implementation Details}\label{sec: implementation}
We have implemented \nm with approximately 3000 lines of Python code and packaged it as a serverless function~\cite{yang2021aid} for ease of use in \cloud. The iPACK system is deployed on a CentOS Linux server with 60GB of RAM and an Intel(R) Core(TM) i7-5930K CPU @ 3.50GHz. The AIN component of \nm is trained and tested with the GPU acceleration of an NVIDIA GeForce GTX TITAN X. We have set the default hyper-parameters of the AIN as $k$=$128$ and $r$=$256$, and the model is trained until its training loss stops decreasing for ten continuous epochs, using an early stopping approach.
As for the comparative solutions, as they are not open-sourced, we have followed the implementation in their respective papers and leveraged well-established libraries to ensure accuracy. For example, we have used AllenNLP~\cite{allennlp} for LinkCM, scikit-learn~\cite{scikit-learn} and gensim~\cite{gensim} for LWE, and HuggingFace~\cite{huggingface} for BERT.

\subsection{Evaluation Metrics}

\textbf{Metrics for evaluating ticket aggregation (RQ1 and RQ3).} 
Given a sequence of tickets, our approach assigns a unique cluster ID, denoted as "incident-{number}" to tickets that are caused by the same incident. Tickets that are not related to a cloud-side issue are marked with the cluster ID "non-incident".
To evaluate the accuracy of our ticket aggregation, we use the widely accepted Rand Index~\cite{rand1971objective, achtert2012evaluation, gates2017impact} for pair-wise comparison in clustering.
We conduct pair-wise comparisons between the ground-truth cluster ID and the predicted cluster ID for all tickets.
The results are used to calculate the following metrics:
True Positives (TP), which are pairs of duplicate tickets correctly predicted to have the same cluster label;
True Negatives (TN), which are pairs of non-duplicate tickets correctly predicted to have different cluster labels;
False Positives (FP), which are pairs of non-duplicate tickets wrongly predicted to have the same cluster label; and
False Negatives (FN), which are pairs of duplicate tickets wrongly predicted to have different cluster labels.
Based on the results, we use the following metrics to evaluate the aggregation results: $precision=\frac{TP}{TP + FP}$, $recall=\frac{TP}{TP + FN}$, and $F1~score= 2 \cdot \frac{precision~\cdot~recall}{precision~+~recall}$.

\textbf{Metrics for evaluating ticket-event correlation (RQ2).} 
The correlation of tickets with an event, referred to as AIN in Section~\ref{sec: correlation}, is a crucial component of \nm. This component generates a ranked list of potential responsible events for a given ticket based on the probability scores (as determined by AIN's output) in descending order.
To assess the accuracy of this step, we use the metric Acc@K (accuracy@K). For each ticket, if the actual ground-truth event appears within the top-K positions of the list, we consider the ticket to be a "hit". The Acc@K metric is calculated as the ratio of the number of hit tickets to the total number of tickets, represented as $Acc@K=\frac{\#~of~hit~tickets}{\#~of~all~tickets}$.
In our evaluation, we consider three values of K (i.e., 1, 2, and 3) and also compute the average of these three metrics to provide a comprehensive assessment.


\subsection{Experimental Details}

\subsubsection{\textbf{RQ1} The Effectiveness of \nm}
In this RQ, we aim to evaluate how accurately \nm can aggregate the duplicate tickets by comparing it with all comparative solutions (Section~\ref{sec: baselines}).
The evaluation is conducted using datasets $\mathcal{A}$, $\mathcal{B}$ and $\mathcal{C}$ and the results are reported in terms of precision, recall, and F1 score. Precision reflects the degree of correctness in the clustering results, while recall represents the completeness of the results. The F1 score is a balance between precision and recall and provides a comprehensive measure of the overall performance of the approach. The results are presented in Table~\ref{tab: clustering_accuracy}. The highest F1 score is emphasized in \textbf{bold}, and the second-best score is \underline{underlined}.

We can make the following observations:
(1) \nm achieves the best F1 score across all three datasets, i.e., 0.935, 0.871, and 0.894, outperforming the second-best methods by 31.2\%, 12.4\% and 18.4\% in dataset $\mathcal{A}$, $\mathcal{B}$ and $\mathcal{C}$, respectively.
(2) Categorization can achieve the highest precision (0.930$\sim$0.943) although its recall is considerably low (0.205$\sim$0.373).
The reason is that the ticket feature  \textit{category} is defined in a fine-grained manner by support engineers in \cloud.
Therefore, it tends to aggressively split the complete set of duplicate tickets into many small groups, leading to a low recall score. 
However, tickets in each such small group share precisely similar semantics as evidenced by the high precision.
(3) iFeedback, LWE, and BERT show lower precision but higher recall than Categorization. 
The reason is that these methods can capture more coarse-grained semantic similarity between tickets.
Consequently, they can generate larger clusters (higher recall) but introduce additional noises (lower precision)
(4) LinkCM can achieve a higher precision among all baseline methods except Categorization. Moreover, after combining with GIP, LinkCM w/ GIP can increase its recall because more tickets are aggregated together through event-event linking.
However, it still under-performs \nm in terms of the overall F1 score because LinkCM cannot correlate a ticket to an event as accurately as \nm does (will show in RQ2). 
For instance, LinkCM may associate a cluster of similar tickets with the wrong event. Therefore, even though related events are linked together, similar tickets are separated into different clusters, resulting in high precision but low recall.

{
\begin{tcolorbox}[breakable,width=\linewidth-2pt,boxrule=0pt,top=1pt, bottom=0pt, left=1pt,right=1pt, colback=gray!20,colframe=gray!20]
\textbf{Answer to RQ1.} 
\nm achieves the best F1 score among all state-of-the-art baselines across three datasets collected from different regions.
\nm slightly sacrifices precision compared with the Categorization method but achieves the highest F1 score 0.871$\sim$0.935, outperforming state-of-the-art methods by 12.4\%$\sim$31.2\%.
\end{tcolorbox}
}

\subsubsection{\textbf{RQ2} The Effectiveness of ticket-event correlation}\label{sec: exp_corr_acc}
In this RQ, the focus is on evaluating the accuracy of the ticket-event correlation step of \nm, i.e., the proposed attentive interaction Network (AIN).
The performance of AIN is compared with LinkCM~\cite{gu2020efficient} and four popular machine learning algorithms: LR (logistic regression), SVM (support vector machine), RF (random forest), and LightGBM (light gradient boosting machine). Additionally, the contribution of the attentive feature interaction component to AIN is studied.

To ensure a fair comparison, categorical features are represented as one-hot vectors, which are then concatenated with the representation of textual features extracted using BERT. This allows for a consistent input feature representation for all models compared. A variant of AIN is also developed by removing its attentive feature interaction component (referred to as "AIN w/o atten." in Table~\ref{tab: corr_results}). This variant instead concatenates all feature embeddings into a single feature vector as the input for the prediction layer, as illustrated in Fig.~\ref{fig: AIN_framework}.
For clarity, this experiment is conducted using all pairs of ticket-event data from datasets $\mathcal{A}$, $\mathcal{B}$ and $\mathcal{C}$.
We compare AIN with the baselines and its variant in terms of Acc@1, Acc@2, Acc@3 and the average of these metrics.

We can make the following observations in the results shown in Table~\ref{tab: corr_results}: 
(1) The proposed AIN model outperforms all baseline models in terms of all four evaluation metrics. Notably, AIN achieves the highest Acc@1 score of 0.817, indicating its superior ability in accurately linking tickets to events and facilitating more effective ticket aggregation.
(2) The introduction of the attentive feature interaction component results in significant improvements in AIN's performance, with a 21.4\% increase in Acc@1 and a 17.8\% increase in the average accuracy. This demonstrates that the component plays a crucial role in identifying effective feature combinations for accurate ticket-event linking.
(3) Interestingly, AIN w/o atten. underperforms LinkCM and achieves similar performance as LightGBM.
The reason is that AIN w/o atten. adopts simple concatenation of feature embedding, which fails to capture effective feature combinations.
(4) LinkCM can outperform other baseline methods since its decomposable attention mechanism is able to capture the semantic matching between tickets and events. 
On the other hand, the relatively low Acc@1 scores of LR, SVM, RF and LightGBM may be due to the sparsity and high dimensionality of the input features. However, RF and LightGBM exhibit improved accuracy over LR and SVM, as they alleviate these problems through feature selection.

\begin{table}[tbp]
\caption{Effectiveness of correlating a ticket to an event.}
\label{tab: corr_results}
\small
\begin{tabular}{c|cccc}
\toprule
Models   & Acc@1                      & Acc@2 & Acc@3 & Average \\ 

\midrule
\midrule
LR	& 0.519 & 0.657 & 0.733 & 0.636 \\
SVM	& 0.332 & 0.409 & 0.493 & 0.411 \\
RF	& 0.563 & 0.684 & 0.761 & 0.669 \\
LightGBM	& 0.658 & 0.723 & 0.832 & 0.712 \\
LinkCM	& 0.743 & 0.769 & 0.882 & 0.798 \\
\midrule
AIN w/o atten.  & 0.673 & 0.762 & 0.824 & 0.753 \\
AIN & \textbf{0.817} & \textbf{0.907} & \textbf{0.936} & \textbf{0.887}   \\ 
 $\Delta$(\%) & +21.4\% & +19.0\% & +13.6\% & +17.8\%  \\
\bottomrule
\end{tabular}
\end{table}

{\begin{tcolorbox}[breakable,width=\linewidth-2pt,boxrule=0pt,top=1pt, bottom=0pt, left=1pt,right=1pt, colback=gray!20,colframe=gray!20]
\textbf{Answer to RQ2.} AIN outperforms all other baseline methods by a large margin in correlating a ticket to the event that causes it. 
The proposed attentive feature combination is the key to achieve the performance, which improves the average accuracy of AIN by 17.8\%.\end{tcolorbox}
}

\subsubsection{\textbf{RQ3} The impact of graph-based incident profiling (GIP) of \nm}\label{sec: GIP_ablation_study}
\begin{figure}[t]
    \centering
    \includegraphics[width=0.9\linewidth]{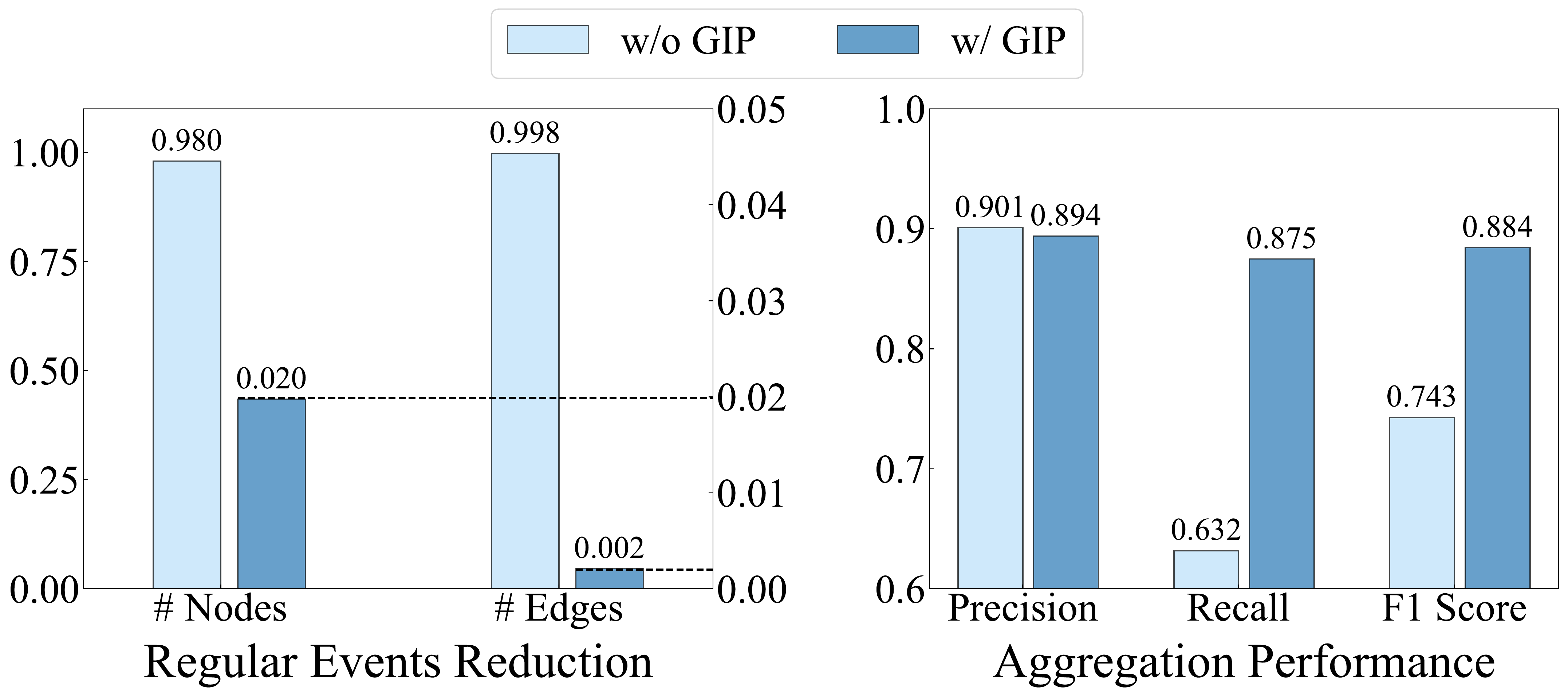}
    \caption{The Effectiveness of Graph-based Profiling (GIP)} 
    \label{fig: GIP_ablation}
\end{figure}

We propose GIP to reduce regular events (noisy events) and link correlated indicative events to profile an incident, which bridges the tickets linked to the events even though they are semantically different.
We evaluate its impact on \nm using the union of all three datasets as in RQ2. 
We conduct the evaluation from the following two aspects:

(1) \textit{The ratio of events reduced.} 
GIP builds a fully-connected event graph (link every two events with positive PMI values) and then prunes this graph via Algorithm~\ref{algo: graph_construction}.
We measure the effectiveness of GIP with the ratio of nodes and edges that are pruned (reduced).
Fig.~\ref{fig: GIP_ablation} (left) presents the ratio of nodes and edges in the event graph without or with GIP (we normalize the ratio for better presentation).
We can observe that only 2\% of nodes and 0.2\% of edges remain after using GIP, which shows GIP can reduce the large volume of events effectively.

(2) \textit{The impact on the overall performance in aggregating duplicate tickets.} 
Though GIP can reduce the number of events, we aim to further evaluate whether it can accurately remove the regular events and link the correlated events as expected.
To achieve this, we compare the ticket aggregation performance of \nm with or without GIP. 
After removing GIP, we regard those tickets linked to the same event by AIN as belonging to the same cluster.
The results are shown in Fig.~\ref{fig: GIP_ablation} (right).
We can observe that after applying GIP, its precision drops slightly, but the recall is largely improved.
As a result, the overall F1 score is improved by 18.9\%, from 0.743 to 0.884.
This indicates that only a small portion of events are not correctly linked; 
however, 
more duplicate tickets are accurately aggregated via event-event linking. 

{
\begin{tcolorbox}[breakable,width=\linewidth-2pt,boxrule=0pt,top=1pt, bottom=0pt, left=1pt,right=1pt, colback=gray!20,colframe=gray!20]
\textbf{Answer to RQ3.} GIP can greatly boost the overall performance of \nm. On the one hand, GIP reserves only 2\% nodes and 0.2\% edges in the pruned event graph. On the other hand, GIP accurately reserves and links the indicative events and improves the F1 score from 0.743 to 0.884.
\end{tcolorbox}
}



\section{Industrial Experience}\label{sec: case_study}

In this section, we share our industrial experience by presenting a success case and a failure case from the real-world deployment of \nm in \cloud. 

\definecolor{crimsonglory}{rgb}{0.75, 0.0, 0.2}
\definecolor{darkblue}{rgb}{0.0, 0.0, 0.55}
\subsection{A success case}
In September 2021, a datacenter maintenance activity resulted in the accidental shutdown of a water tower pump, which is a critical component of cooling systems. To prevent overheating and potential damage to users' data, the maintenance personnel had to shut down the downstream storage hardware. This caused a storage service disruption, leading to cascading impacts on several dependent services such as the SQL DB and Workflow App, and triggering alerts.

\begin{figure}[t]
    \centering
    \includegraphics[width=0.85\linewidth]{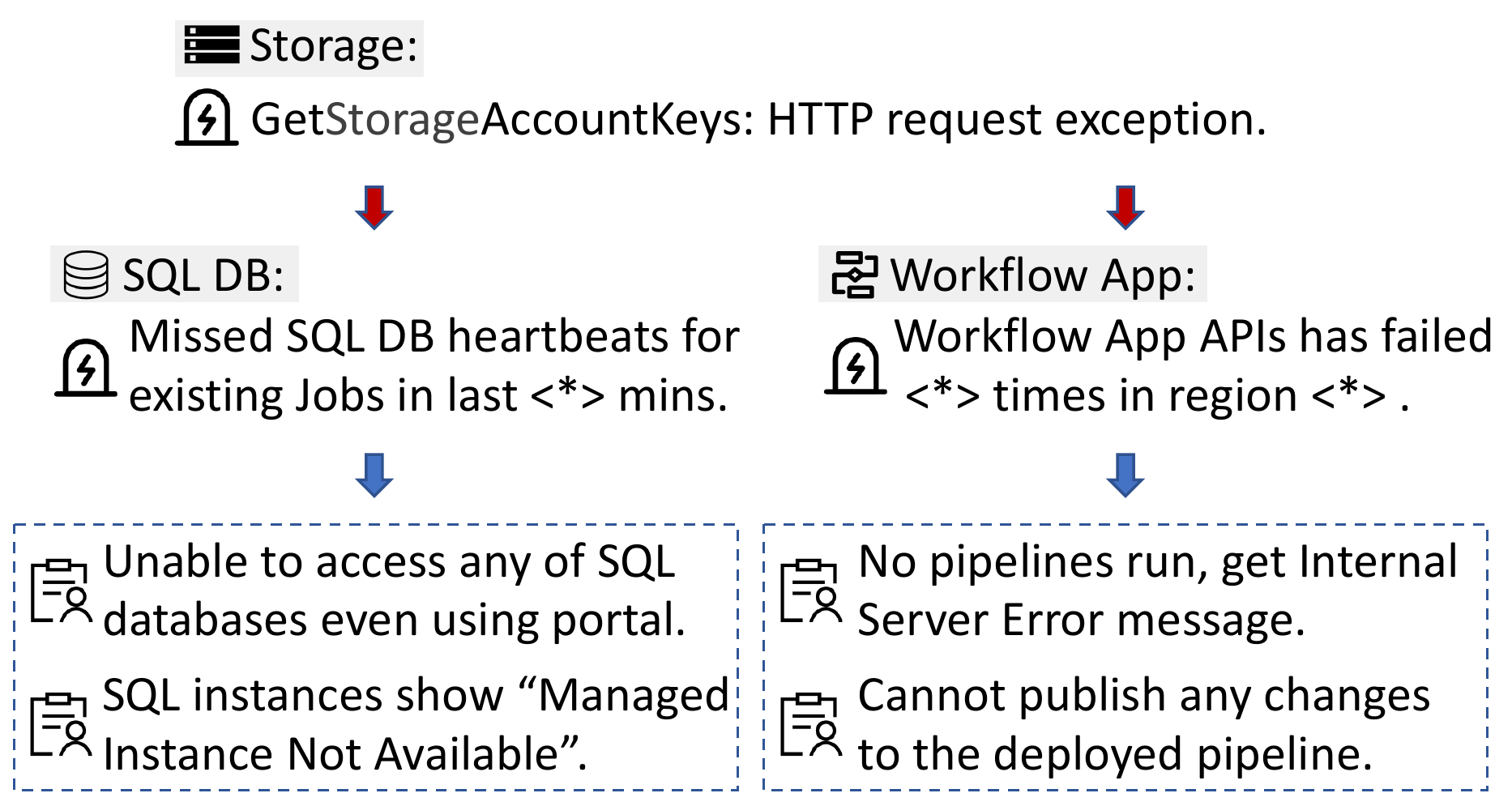}
    \caption{A success case of \nm in \cloud} 
    \label{fig: success_case}
\end{figure}

The CSS team received a substantial number of tickets describing a wide range of issues in response to these events. To assist with the situation, \nm continuously collected and analyzed the generated alerts and tickets. The partial output of \nm's analysis is presented in Fig.~\ref{fig: success_case}.
\nm successfully linked the storage alert with corresponding alerts from SQL DB and Workflow App, as demonstrated by the \textcolor{crimsonglory}{\textbf{red}} arrows in Fig.~\ref{fig: success_case}. Additionally, the tickets caused by these events were linked to their respective root cause events, as depicted by the \textcolor{darkblue}{\textbf{blue}} arrows. This allowed the tickets to be aggregated, despite their semantic differences, and the results were pushed to the support engineers.
With the information provided by \nm, support engineers were able to initiate batch communications with potentially impacted customers and avoid duplicative manual inspections. Throughout the resolution process, the customers were continuously informed of the mitigation progress of the incident.


\subsection{A failure case}
\nm could sometimes fail when it cannot find responsible alerts in the cloud systems for a ticket. In August 2021, the CSS team received multiple tickets complaining of 503 (service unavailable) errors when the customers were using Web Services. Though the tickets were suspected to be caused by an internal issue due to their similar symptoms, \nm did not correlate them with any alert. Only around five hours after the first ticket had been received, a related alert was fired and correlated by \nm. 
According to the after-the-fact analysis of on-call engineers, the root cause of this incident turned out to be bad configurations of a Canary (gray) release for a few tenants. 
The developers did not configure a specific monitor for each of the tenants but monitored all tenants as a whole.
As a result, the monitor was not sensitive enough and only triggered when most of the tenants' requests failed. 
Nevertheless, \nm continuously runs and could still correlate the alert with the resultant tickets after the alert was finally fired. In this way, \nm can potentially discover such under-monitoring cases and guide the configuration of monitors to improve system reliability~\cite{li2022intelligent}.
Fortunately, such cases (tickets submissions before alerts) are rare in \cloud with comprehensive monitoring according to our study (Section~\ref{sec: alert_ticket_relation}).

\section{Related Work}

\subsection{Incident Analysis}
Researchers have devoted sustained efforts on empirical studies~\cite{gunawi2016does, huang2017gray, liu2019bugs, cotroneo2019bad, chen2020towards} of cloud incidents in the last few years. Gunawi et al.~\cite{gunawi2016does} discussed why outages still take place in cloud environments by analysing headline news and public postmortem reports of 32 popular Internet services. Huang et al.~\cite{huang2017gray} discussed their experiences with gray failure in production cloud-scale systems and demonstrated its broad scope and consequences. Chen et al.~\cite{chen2020towards} presented a comprehensive study on how alerts and incidents are managed in large-scale public cloud. 
Cloud alerts are notoriously blamed for its great volume. 
In general, there are two threads of studies proposed towards resolving the challenge. The major thread aims to correlate alerts that are caused by the same incident~\cite{wang2021fast}\cite{gu2020efficient}\cite{zhao2020understanding}. 
Given a large number of alerts happening, Chen et al.~\cite{chen2020incidental} empirically found that only a small portion of alerts matters and proposed to prioritize alerts based on historical data. 
Chen et al.~\cite{chen2020identifying}\cite{chen2019outage} proposed to predict the link between two alerts by combining alert textual information and the topology information among alerts (i.e., the topology of components that generate these alerts). 
These studies either require experts' manual annotations~\cite{chen2022online}\cite{chen2020identifying} or precise system topology~\cite{zhao2020understanding}\cite{chen2021graph}. Differently, we propose GIP, which does not require such labels or prior knowledge to identify alert-alert relations. We further leverage the alert-alert relations to aggregate tickets for efficient processing and management.


\subsection{Issue Report Analysis}
Issue reports, including app reviews, user feedback, bug reports, test reports, GitHub issues, support tickets, etc., are crucial for service providers to gain a better understanding of their customers' experiences. 
A large body of research has been devoted to the analysis of issue reports, covering topics such as duplicate bug reports detection~\cite{wang2008approach}\cite{ nguyen2012duplicate}\cite{zhou2012learning}\cite{ budhiraja2018lwe}, emerging issue detection~\cite{gao2018online} \cite{zheng2019ifeedback}\cite{gao2019emerging} bug reproduction~\cite{zhao2019recdroid}\cite{cao2014symcrash}, bug report summarization~\cite{rastkar2014automatic}\cite{li2018unsupervised} and empirical studies~\cite{kucuk2021characterizing}\cite{zou2018practitioners}\cite{ma2017developers}.

Most existing studies focus on natural language text information such as titles and descriptions. In addition, some latest attempts~\cite{he2020duplicate}\cite{liu2020clustering}\cite{cooper2021takes} proposed to jointly consider multi-modality features, e.g., text and images (e.g., app screenshots), which has become a recent hot trend in the research direction.
Different from these studies that purely focus on the customer-side issue report information, in this work, we also consider ongoing alerts and incidents in the complex cloud system. We aim to bridge the cloud alerts with cloud users' tickets to facilitate efficient ticket processing.

\section{Threats to Validity}


\textbf{External Validity.} The study's object is the primary external threat. The data was collected from \cloud, as there is no publicly available dataset containing customer tickets and a large number of alerts. However, \cloud is a world-leading cloud provider with a vast scale. The data covers a broad range of services from various regions (Section~\ref{sec: dataset}). Hence, the evaluation in \cloud should be representative and convincing. Furthermore, \nm leverages the common features provided by the most popular cloud vendors (Section~\ref{sec: bg_alert_and_tickets}), making it capable of generalizing to similar cloud systems, potentially benefiting cloud customers globally.

\textbf{Internal Validity.} Implementation and parameter setting are the main internal threats to validity. For implementation, the baseline approaches are not open-sourced, so we re-implemented them by following the original papers closely. To reduce the implementation threat, we leveraged several mature libraries for implementing the core algorithms (Section~\ref{sec: baselines}). Both the proposed and baseline methods underwent peer code review. For parameter setting, we tuned all methods through grid-search and chose the best results.

\section{conclusion}

This paper tackles the problem of aggregating duplicate customer support tickets for cloud systems.
Previous solutions that mainly rely on customer-side information (i.e., textual similarity between tickets) are sub-optimal for tickets of large-scale cloud systems. 
The main cause is the complexity of cloud systems that consist of many inter-dependent services, where the customers may experience distinct issues even though they are affected by the same incident.
To overcome this limitation, 
we propose \nm to leverage alerts of cloud systems to facilitate ticket aggregation. 
Specifically, we propose graph-based incident profiling (GIP) to model alert-alert relations and attentive interaction network (AIN) to model alert-ticket relations, respectively.
In this way, we can aggregate the tickets that are linked to the same incident (linked alerts) even though they carry dissimilar semantics.
We evaluate \nm based on three datasets collected from the real-world production environment in a large-scale cloud vendor, \cloud.
\nm achieves the F1 score of 0.871$\sim$0.935 and outperforms state-of-the-art methods by 12.4\%$\sim$31.2\% across the three datasets.

For future work, we will deploy \nm to more services in \cloud and conduct rigorous user studies among the support engineers to understand the usefulness in accelerating support tickets. In addition, we plan to extend \nm with the ability to conduct root cause analysis based on the correlated alerts.


\section{Data Availability}
The ticket data used in this work is collected from a real-world cloud vendor, which is highly confidential and contains a lot of personally identifiable information (PII). 
To protect customers' privacy, we decide not to release the original dataset. However, to facilitate the community to benefit from our work, we release the source code of \nm together with some \textit{synthetic data samples} on Github (\url{https://github.com/OpsPAI/iPACK.git}).

\section{Acknowledgement}\label{sec:Acknowledgement}

The work described in this paper was supported by the Research Grants Council
of the Hong Kong Special Administrative Region, China (No. CUHK 14206921 of the
General Research Fund) and Australian Research Council (ARC) Discovery Projects (DP200102940, DP220103044).

\balance
\bibliographystyle{IEEEtran}
\bibliography{icse23}

\end{document}